# Influence of interstitial Fe to the phase diagram of Fe$_{1+y}$Te$_{1-x}$Se$_x$ single crystals

Yue Sun[*], Tatsuhiro Yamada, Sunseng Pyon, Tsuyoshi Tamegai

*Department of Applied Physics, The University of Tokyo, 7-3-1 Hongo, Bunkyo-ku, Tokyo 113-8656, Japan*

Email addresses: [*]sunyue.seu@gmail.com

## *Abstract*

Superconductivity (SC) with the suppression of long-range antiferromagnetic (AFM) order is observed in the parent compounds of both iron-based and cuprate superconductors. The AFM wave vectors are bicollinear ($\pi$, 0) in the parent compound FeTe different from the collinear AFM order ($\pi$, $\pi$) in most iron pnictides. Study of the phase diagram of Fe$_{1+y}$Te$_{1-x}$Se$_x$ is the most direct way to investigate the competition between bicollinear AFM and SC. However, presence of interstitial Fe affects both magnetism and SC of Fe$_{1+y}$Te$_{1-x}$Se$_x$, which hinders the establishment of the real phase diagram. Here, we report the comparison of doping-temperature ($x$-$T$) phase diagrams for Fe$_{1+y}$Te$_{1-x}$Se$_x$ ($0 \leq x \leq 0.43$) single crystals before and after removing interstitial Fe. Without interstitial Fe, the AFM state survives only for $x < 0.05$, and bulk SC emerges from $x = 0.05$, and does not coexist with the AFM state. The previously reported spin glass state, and the coexistence of AFM and SC may be originated from the effect of the interstitial Fe. The phase diagram of Fe$_{1+y}$Te$_{1-x}$Se$_x$ is found to be similar to the case of the "1111" system such as LaFeAsO$_{1-x}$F$_x$, and is different from that of the "122" system.

The discovery of superconductivity (SC) in iron-based superconductors (IBSs)[1] provides another route to realize SC at high temperatures other than the cuprates. Some similarities between IBSs and cuprates give us important clues to the understanding of the mechanism of high temperature SC. Among those similarities like layered structure and very high critical field[2], the most important aspect is that both systems maintain a long-range antiferromagnetic (AFM) order in the parent compounds, and the SC emerges after the suppression of the AFM order[3]. Thus, the study of phase diagram becomes the most direct way to investigate the relation between AFM and SC. Until now, integrated phase diagrams for some members of IBSs have already been well established, especially in the "122" system since single crystals with sufficient dimensions and good quality can be obtained easily[4]. Some interesting phenomena are observed like the coexistence of AFM and SC in under-doped region[4], asymmetric superconducting dome in $Ba_{1-x}K_xFe_2As_2$[5], nematic orders[6] and quantum critical point (QCP)[7]. All these discoveries in the past several years have promoted our understanding of the mechanism of SC in IBSs. Spin fluctuations related to the nesting of disconnected electron and hole Fermi surfaces[8], are proposed to be responsible for the high value of $T_c$ in IBSs based on the $s_\pm$ scenario[9]. In addition, the contribution of large orbital fluctuation has also been stressed from the $s_{++}$ scenario[10].

On the other hand, iron chalcogenides recently attracted much more attention in IBSs because of its unexpected high $T_c$. Although the initial $T_c$ in FeSe was only 8 K[11], it increased up to 14 K[12] with appropriate Te substitution and up to 37 K[13,14] under high pressure. Furthermore, by intercalating spacer layers between adjacent FeSe layers, $T_c$ has reached ~32 K[15] in $A_xFe_{2-y}Se_2$ (A=K, Cs, Rb and Tl) and 43 K[16] in $Li_x(NH_2)_y(NH_3)_{1-y}Fe_2Se_2$ ($x \sim 0.6$; $y \sim 0.2$). By applying pressure to $A_xFe_{2-y}Se_2$, $T_c$ can even reach ~48 K[17]. Furthermore, the monolayer of FeSe grown on $SrTiO_3$ even shows a sign of SC over 100 K[18]. Among iron chalcogenides, $Fe_{1+y}Te_{1-x}Se_x$ is unique in its structural simplicity, consisting of only iron-chalcogenide layers, which is ideal for probing the mechanism of SC. Although $Fe_{1+y}Te_{1-x}Se_x$ shows some similarities to iron pnictides like the Fermi surface topology which is characterized by hole bands around $\Gamma$ point and electron bands around $M$ point[8], it manifests some unique properties different from iron pnictides. The most crucial one is the antiferromagnetic wave vectors, which is bicollinear ($\pi$, 0) in the parent compound FeTe[19] different from the collinear antiferromagnetic order ($\pi$, $\pi$) in most of iron pnictides[20]. Since the AFM order is believed to be related to the high temperature SC, a systematic study of the competition between bicollinear AFM and SC orders with doping is crucial to the understanding of its paring mechanism. Furthermore, the phase diagram of $Fe_{1+y}Te_{1-x}Se_x$ will give us another opportunity to testify some phenomena observed in iron pnictides like the coexistence of AFM and SC, and the possible QCP.

Until now, although several phase diagrams have been already reported based on $Fe_{1+y}Te_{1-x}Se_x$ single crystals[21-25] and even thin films[26,27], they are all under debate, especially in the low Se doping region. Some basic information is even controversial in those reported results, like the region of bulk SC, the coexistence of AFM and SC, and the spin glass state. These controversies are believed to come from the sample-dependent Fe nonstoichiometries,[19,28] which originate from the partial occupation of the second Fe site (interstitial Fe site) in the Te/Se layer. The interstitial Fe with valence near $Fe^+$ will provide an electron into the 11 system[29]. The interstitial Fe is also strongly magnetic, which provides local moments that interact with the adjacent Fe layers[29]. In the parent compound $Fe_{1+y}Te$, the long-range ($\pi$, 0) order can be tuned from commensurate to incommensurate by changing the amount of interstitial Fe[19]. Furthermore, the magnetic moment from interstitial Fe will act as a pair breaker and also localize the charge carriers[30,31]. Thus, the existence of interstitial Fe, which is easily formed in the standard growth technique employing slow cooling and their amount varies among different groups[32], makes the phase diagram of $Fe_{1+y}Te_{1-x}Se_x$ still unclear until now.

Recently, our $O_2$-annealing technique with fine tuning capability was proved to be very effective in minimizing the detrimental effect of the interstitial Fe and including bulk SC with a large value of normalized specific heat jump at $T_c$[33]. In this report, we adopt the $O_2$-annealing technique to $Fe_{1+y}Te_{1-x}Se_x$ single crystals with doping level $0 \leq x \leq 0.43$ to minimize the effect of the interstitial Fe. The doping-temperature ($x$-$T$) phase diagrams for $Fe_{1+y}Te_{1-x}Se_x$ ($0 \leq x \leq 0.43$) single crystals before and after removing interstitial Fe were established and compared based on the systematic studies of the structure, magnetic, and transport properties. Results show that the phase diagram is largely affected by the amount of interstitial Fe for all the doping levels. Without interstitial Fe, the AFM state is found to survive only in a narrow region of $x < 0.05$, and bulk SC emerges from $x =$ 0.05, and does not coexist with the AFM state. The previously reported spin glass state, and the coexistence of AFM and SC may be originated from the effect of interstitial Fe. The phase diagram of FeTe$_{1-x}$Se$_x$ after removing the interstitial Fe is found to be similar to the case of the "1111" system such as LaFeAsO$_{1-x}$F$_x$[34], and is different from that of the "122" system.

**Results**

Figure 1(a) shows the single crystal XRD patterns for the as-grown Fe$_{1+y}$Te$_{1-x}$Se$_x$ ($0 \leq x \leq 0.43$) single crystals. Here, the selenium content $x$ is the analyzed value for a similar piece of crystal taken from the same batch by the inductively-coupled plasma (ICP) atomic emission spectroscopy measurements. Only the (00$l$) peaks are observed, suggesting that the crystallographic $c$-axis is perfectly perpendicular to the plane of the single crystal. With increasing Se doping, the positions of (00$l$) peaks gradually shift to higher values of 2$\theta$. The lattice constant $c$ is calculated and plotted in Figure 1(c), which is almost linearly decreasing with increasing Se doping similar to that reported in a previous report[35]. After removing the interstitial Fe by O$_2$-annealing, the positions for (00$l$) peaks change little, as shown in Figure 1(b) for a typical example of (003) peaks for Fe$_{1+y}$Te$_{0.57}$Se$_{0.43}$ before and after annealing. The lattice constant $c$ for the annealed crystals is also plotted and compared in Figure 1(c), which shows that the interstitial Fe affects little to the $c$-axis lattice constant. Actually, previous analyses proved that the lattice constant $a/b$ is slightly decreased after removing the interstitial Fe, although the lattice constant $c$ changes little[36].

To probe the influence of Se doping to the SC in Fe$_{1+y}$Te$_{1-x}$Se$_x$, temperature dependence of zero-field-cooled (ZFC) and field-cooled (FC) magnetization at 5 Oe were measured for the as-grown and annealed crystals. All the as-grown crystals usually show no SC or very weak diamagnetic signal. After removing the interstitial Fe by annealing, SC emerges from $x = 0.05$, and the value of $T_c$ is gradually enhanced with the increase of Se doping up to 14.5 K in Fe$_{1+y}$Te$_{0.57}$Se$_{0.43}$ as shown in Figure 2. Besides, all the annealed crystals show relatively sharp SC transition width $\Delta T_c \leq 1$ K. The SC observed in the annealed crystals has already been proved to be in bulk nature by the clear specific heat jump and a large value of critical current density, $J_c$, in our previous report[33]. Actually, when the Se doping level is equal or larger than 0.05, all the annealed crystals show large value of $J_c$ ~$3 \times 10^5$ A/cm$^2$ at 2 K under self-field similar to that reported for the crystal with $x = 0.43$ [37,38].

Figure 3(a) and (b) show the normalized magnetic susceptibilities measured under 10 kOe magnetic field parallel to $c$-axis for the as-grown and annealed Fe$_{1+y}$Te$_{1-x}$Se$_x$ ($0 \leq x \leq 0.43$) single crystals, respectively. It is obviously that the as-grown FeTe shows a sharp transition at ~58 K, which is due to the antiferromagnetic (AFM) transition based on the previous report[23]. With Se doping, the AFM transition temperature $T_N$ is gradually suppressed to lower temperatures, and becomes much broader at $x = 0.09$. After that, the AFM transition disappears and is replaced by a very broad hump-like feature. Such a hump-like feature may be originated from the spin glass state according to the neutron scattering results[21]. The hump-like feature survives up to $x = 0.33$, and is not observed for $x \geq 0.43$.

In crystals after annealing, the value of magnetic susceptibility does not show a systematic evolution and is irregular, which is caused by the magnetism from some Fe impurities. During the annealing process, the interstitial Fe are removed from their original positions (interstitial sites in Te/Se layers), and form some compounds like Fe$_2$O$_3$ or FeTe$_2$[33,39,40]. Although those impurities are mainly formed in the surface layers, and removed by polishing before measurements, small parts may still remain inside the crystals and disturb the magnetic susceptibility value because of their strong magnetism. However, we can still obtain some important information from the data regardless of the irregularity in the absolute value. As shown clearly in Figure 3(b), the value of $T_N$ for the pure FeTe is enhanced to ~72 K after removing the interstitial Fe. The AFM transition is only observed in crystals with $x = 0$ and 0.03. When the Se doping level increases over 0.05, the AFM is totally suppressed. On the other hand, the hump-like feature observed in the as-grown crystals is not witnessed after annealing. For $x > 0.03$, the annealed crystals only show the SC transition at low temperatures.

Figure 4 shows the temperature dependence of the in-plane resistivity $\rho$ ($T$) for the as-grown and annealed Fe$_{1+y}$Te$_{1-x}$Se$_x$ ($0 \leq x \leq 0.43$) single crystals. For the as-grown crystals, the AFM transition can be observed in the doping region of $0 \leq x \leq 0.05$ as indicated by the solid magenta arrows. The values of $T_N$ are close to those obtained from magnetic susceptibility measurements. For $x \geq 0.05$, the SC transition can be observed and indicated by the dashed blue arrows. However, the SC can be only observed in the resistivity measurements. Neither the diamagnetic signal nor the jump at $T_c$ in specific heat can be observed, which indicates that the SC observed here are filamentary in nature[33]. Furthermore, temperature dependence of resistivity for all the as-grown crystals manifests a nonmetallic behavior ($d\rho/dT < 0$) with decreasing temperature below 150 K. Such nonmetallic resistivity behavior is caused by the localization effect from interstitial Fe[30,41], which is suppressed and replaced by a metallic behavior ($d\rho/dT > 0$) after removing the interstitial Fe by $O_2$-annealing as shown in the right panel of Figure 4. For the annealed crystals, the AFM transition, marked by the solid magenta arrows, can be observed only in the doping region of $x \leq 0.03$, which is consistent with the results of magnetic susceptibility. For $x \geq 0.05$, the SC transition can be observed. Since the SC observed here is bulk in nature as discussed before, the positions of $T_c$ are indicated by using solid blue arrows. It is clear that the value of $T_c$ gradually increases with the Se doping. Here, we should point out that a SC-like transition at low temperature is observed in the annealed crystal with $x = 0.03$, however, the zero resistivity is not reached in the measured low temperature limit of 2 K. Such a SC transition is filamentary in nature, since is not observed in magnetization measurements. It may come from the atomic-size fluctuation of Se doping or possible local strain effect.

To get more insight into the influence of interstitial Fe to the transport properties, temperature dependence of the Hall coefficients, $R_H$, for the as-grown and annealed $Fe_{1+y}Te_{1-x}Se_x$ ($0 \leq x \leq 0.43$) single crystals are measured and shown in Figure 5. For the as-grown crystals, obvious AFM transition can be observed in Se doping region of $0 \leq x \leq 0.09$, and the transition temperatures $T_N$ are indicated by the solid magenta arrows, which is consistent with the magnetic susceptibility results. For the as-grown crystal with $x = 0.09$, the AFM transition becomes much broader. Such broader transition is also witnessed in the magnetic susceptibility measurement, which indicates that $x = 0.09$ is close to the edge of the AFM region. Since the AFM in the crystal with $x = 0.09$ is already very weak, it is not observed in the temperature dependence of resistivity measurements. The $R_H$ for the as-grown crystals all show positive values before the AFM transition, which indicates that the hole-typed charge carriers are dominant. Besides, for $x > 0.09$, $R_H$ shows an obvious upturn behavior with decreasing temperature below 100 K. Such upturn behavior can be also explained by the localization effect due to the presence of the interstitial Fe[30,31].

For the annealed crystals, AFM transition is only observed in the crystals with $x = 0$ and 0.03, and the value of $T_N$ for FeTe is increased after annealing, which are all consistent with both the magnetic susceptibility and temperature dependence of resistivity measurements. For $x \geq 0.05$ (the bulk SC region), $R_H$ keeps nearly temperature independent behavior at high temperatures, followed by a slight increase below 100 K, then suddenly decreases before reaching $T_c$. The value even changes sign from positive to negative in the crystal with $x = 0.43$. The characteristic temperature at which $R_H$ shows the maximum value before decreasing is defined as $T^*$, and also indicated by the solid orange arrows in the figure. It is obviously that the value of $T^*$ gradually increases with the increase in the Se doping level. The strong temperature-dependent $R_H$ is usually explained by the multiband nature of the sample. For the annealed $Fe_{1+y}Te_{1-x}Se_x$, hole-type charge carriers are dominant at temperatures above ~150 K, since the $R_H$ keeps temperature independent positive value. Below ~150 K, the slight increase in $R_H$ may come from the mobility change of the hole-type carriers or the remaining small amount of impurities formed during the removing process of the interstitial Fe (for example the $FeTe_m$[39]). Below $T^*$, $R_H$ decreases with decreasing temperature, even changes sign to negative for $x = 0.43$, which indicates that the electron-type charge carriers become more dominant. Here, we should emphasize that the $T^*$ shows a coincident behavior with bulk $T_c$ ($T^*$ is observed only in the region of bulk SC, and also increased with Se doping), which indicates that the multiband nature is strongly related to the SC in $Fe_{1+y}Te_{1-x}Se_x$, and is covered up by the effect of interstitial Fe in the as-grown crystals. Actually, a very broad hump-like behavior can be observed in the $\rho(T)$ curves for all the crystals with bulk SC, which may have similar origination as the $T^*$ in Hall effect.

## Discussion

Based on the magnetization, magnetic susceptibility, resistivity, and Hall effect measurements described above, we can establish a doping-temperature ($x$-$T$) phase diagram for the as-grown and annealed $Fe_{1+y}Te_{1-x}Se_x$ ($0 \leq x \leq 0.43$) single crystals as shown and compared in Figures 6(a) and (b), respectively. For the as-grown crystals, in the doping region of $x < 0.12$, the AFM transition, ~58 K in non-doped FeTe, is monotonically suppressed with increasing Se substitution. More specifically in Figure 6(a), the downtriangles, uptriangles and diamonds represent the Neel temperature $T_N$ obtained by magnetic susceptibility, resistivity and Hall coefficient measurements. And the three disparate measurements are roughly consistent with each other. Accompanied by the suppression of AFM, SC emerges from $x = 0.05$, and coexists with the antiferromagnetic phase until $x < 0.13$. That SC, marked by the squares, is *not* bulk in nature, and can be only observed in resistive measurement. For $x \geq 0.12$, the AFM transition is absent and replaced by a spin glass state (observed by magnetic susceptibility measurements, and marked by the righttriangles). The spin glass state is originated from the effect of interstitial Fe, which interacts with more than 50 neighboring Fe in the adjacent Fe layers, and induces the magnetic Friedel-like oscillation at $(\pi,0)$ order[42].

After removing the interstitial Fe by $O_2$-annealing, the phase diagram of $Fe_{1+y}Te_{1-x}Se_x$ ($0 \leq x \leq 0.43$) is dramatically changed. As shown in Figure 6(b), the AFM state is suppressed into a very narrow region of $x < 0.05$, and the spin glass state is totally suppressed. Immediately after the disappearance of AFM state, bulk SC emerges, and is observed in the doping region of $x \geq 0.05$. The coexistence of AFM and SC states is absent in the annealed crystals. Thus, the previously reported coexistence of AFM, spin glass state with SC may be originated from the effect of interstitial Fe. Besides, the characteristic temperature $T^*$ observed in the $R_H$ is plotted in the figure, which also resides in the doping region of $x \geq 0.05$, and gradually increases with increasing Se doping. It suggests that the multiband effect in $Fe_{1+y}Te_{1-x}Se_x$ may be strongly related to the occurrence of SC. On the other hand, the rapid suppression of AFM state with small amount of doping, absence of coexistence of the AFM and SC states are all similar to the phase diagrams of $LaFeAsO_{1-x}F_x$[34] and $CeFeAsO_{1-x}F_x$[43]. This behavior is quite different from the phase diagram of "122" system[4], where the coexistence of AFM and SC is commonly observed. And the step-like behavior of the magnetism and SC in the small region of $0.03 < x < 0.05$ suggests that the SC in the $Fe_{1+y}Te_{1-x}Se_x$ system may be related to the suppression of static magnetic order rather than the increase of the effective charge carrier density by the doping or structural distortion.

In summary, the doping-temperature ($x$-$T$) phase diagrams for $Fe_{1+y}Te_{1-x}Se_x$ ($0 \leq x \leq 0.43$) single crystals before and after removing interstitial Fe by $O_2$-annealing are established and compared based on the systematical studies of the structure, magnetic, and transport properties. Results show that the phase diagram is largely affected by the interstitial Fe. Without interstitial Fe, the AFM state is found to be suppressed quickly with Se doping, surviving only in a narrow region of $x < 0.05$. The AFM state is proved not to coexist with the bulk SC. The previously reported spin glass state, and the coexistence of AFM and SC may be originated from the effect of interstitial Fe. Besides, a characteristic temperature $T^*$ observed in the temperature dependent Hall coefficient in the annealed crystals is found to be accompanied by the bulk SC, which may indicate the important role of the multi-band effect in the realization of SC in this system. Future efforts on this point may be helpful to the understanding of the paring mechanism of this system.

**Methods**

**Sample growth and annealing.** Single crystals $Fe_{1+y}Te_{1-x}Se_x$ ($0 \leq x \leq 0.43$) are grown by the self-flux method as described in detail elsewhere[41]. Single crystals with Se doping level larger than 43% cannot be grown by the flux method because of the phase separation[44]. All the crystals show plate-like morphology, with $c$-axis perpendicular to the plane of the plate, and can grow up to centimeter-scale. The Se/Te ratio is evaluated by the inductively-coupled plasma (ICP) atomic emission spectroscopy, and its fluctuation in different pieces obtained from the same batch is almost negligible ($\leq 1\%$). In addition, Se/Te ratio is found to change little after annealing ($\leq 1\%$). The energy dispersive x-ray spectroscopy (EDX) measurements show that Te and Se are almost homogeneously distributed in both the as-grown and annealed crystals[38]. The amount of interstitial Fe in the as-grown crystals is estimated as ~10 – 14% based on the ICP results. The obtained as-grown single crystals are then cut and cleaved into thin slices, and annealed with appropriate amount of $O_2$ at 400 °C to remove the interstitial Fe. Details about the $O_2$-annealing processes are reported in our previous publication[33]. Although the interstitial Fe was removed from its initial position by annealing, it may still remain in the crystal, mainly on the surface, in some form of oxides and other binary compounds. Thus, traditional compositional analysis methods like ICP, EDX and electron probe microanalyzer (EPMA) can hardly detect the change of interstitial Fe.

Actually our ICP analyses on the $O_2$-annealed crystal show a small reduction in the Fe content after annealing. To observe the change in the amount of interstitial Fe, we employ the scanning tunneling microscopy (STM) measurements, which can directly observe the interstitial Fe in Te/Se layers. Based on our previous result on the crystal with $x = 0.43$, the values of $T_c$ and $J_c$ are gradually increased with removing the interstitial Fe, and reach the maximum values when the interstitial Fe are almost totally removed as observed in the STM images[33]. In the current research, all the crystals used as the annealed ones are carefully annealed by the same method to the stage with maximum $T_c$ and $J_c$, which are believed to contain little interstitial Fe.

**Measurements and verifications.** Details of the lattice constant change by the annealing process is characterized by means of X-ray diffraction with Cu-$K\alpha$ radiation. Magnetization measurements are performed to check the superconducting transition temperature $T_c$, and the susceptibility by using a commercial superconducting quantum interference device (SQUID). Longitudinal and transverse (Hall) resistivity measurements are performed by the six-lead method with a Quantum Design physical property measurement system (PPMS). In order to decrease the contact resistance, we sputter gold on the contact pads just after the cleavage. Then gold wires are attached on the contacts with silver paste. The Hall resistivity $\rho_{yx}$ is extracted from the difference of the transverse resistance measured at positive and negative fields, i.e., $\rho_{yx}(H) = [\rho_{yx}(+H) - \rho_{yx}(-H)]/2$, which can effectively eliminate the longitudinal resistivity component due to the misalignment of contacts. Hall coefficients $R_H$ is estimated from $R_H = \rho_{yx}/\mu_0 H$.

## Acknowledgements

Y.S. gratefully appreciates the support from Japan Society for the Promotion of Science.

## Author contributions

Y.S performed most of the experiments and analyzed the data. T.Y and S.P contributed to the XRD measurement. Y.S, and T.T designed the research. Most of the text of the paper was written jointly by Y.S and T.T. All the authors contributed to discussion on the results for the manuscript.

## Author Information

The authors declare that they have no competing financial interests. Correspondence and requests for materials should be addressed to Y.S ([sunyue.seu@gmail.com](sunyue.seu@gmail.com))

# Figure captions

Figure 1: (a) Single crystal X-ray diffraction patterns of the as-grown $Fe_{1+y}Te_{1-x}Se_x$ ($0 \leq x \leq 0.43$) single crystals. (b) Comparison of the (003) peaks for the as-grown and $O_2$-annealed $Fe_{1+y}Te_{0.57}Se_{0.43}$. (c) Lattice constant $c$ for $Fe_{1+y}Te_{1-x}Se_x$ before and after annealing.

Figure 2: Temperature dependence of zero-field-cooled (ZFC) and field-cooled (FC) magnetization at 5 Oe for the $O_2$-annealed $Fe_{1+y}Te_{1-x}Se_x$ ($0.05 \leq x \leq 0.43$) single crystals.

Figure 3: Magnetic susceptibilities measured at 10 kOe with $H \parallel c$ for $Fe_{1+y}Te_{1-x}Se_x$ ($0 \leq x \leq 0.43$) (a) before and (b) after $O_2$-annealing.

Figure 4: Temperature dependence of in-plane resistivity for the as-grown (left panel) and $O_2$-annealed (right panel) $Fe_{1+y}Te_{1-x}Se_x$ ($0 \leq x \leq 0.43$) single crystals. The solid magenta arrows, dashed blue arrows and solid blue arrows are used to mark the AFM, non-bulk SC and bulk SC transitions, respectively.

Figure 5: Temperature dependence of Hall coefficients for the as-grown (left panel) and $O_2$-annealed (right panel) $Fe_{1+y}Te_{1-x}Se_x$ ($0 \leq x \leq 0.43$) single crystals. The AFM transition temperatures $T_N$ and characteristic temperature $T^*$ were marked by the magenta and orange arrows, respectively.

Figure 6: The doping-temperature ($x$-$T$) phase diagrams for $Fe_{1+y}Te_{1-x}Se_x$ ($0 \leq x \leq 0.43$) single crystals (a) before and (b) after $O_2$-annealing obtained from magnetization, magnetic susceptibility, resistivity, and Hall effect measurements.

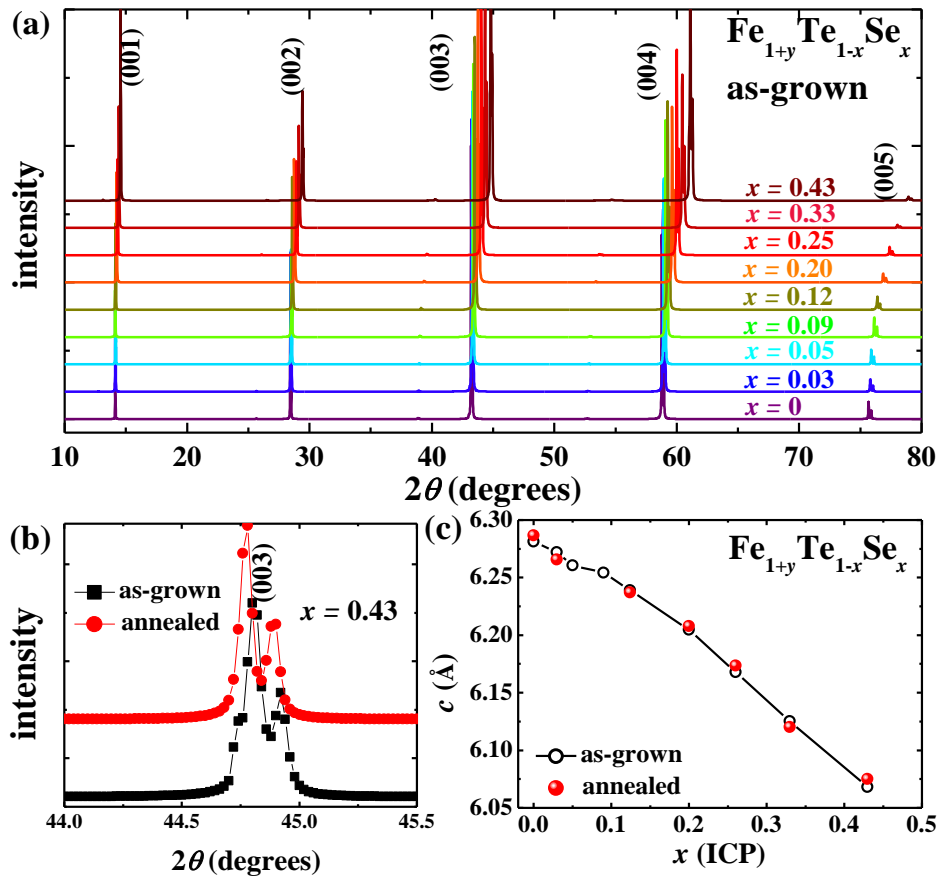

Figure 1

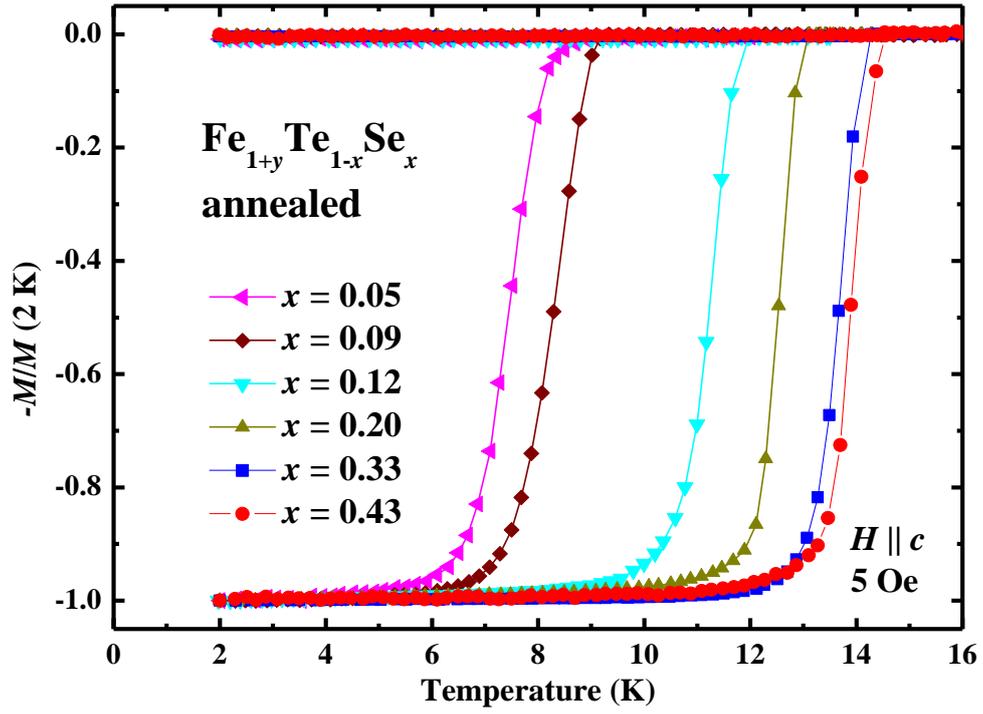

**Figure 2**

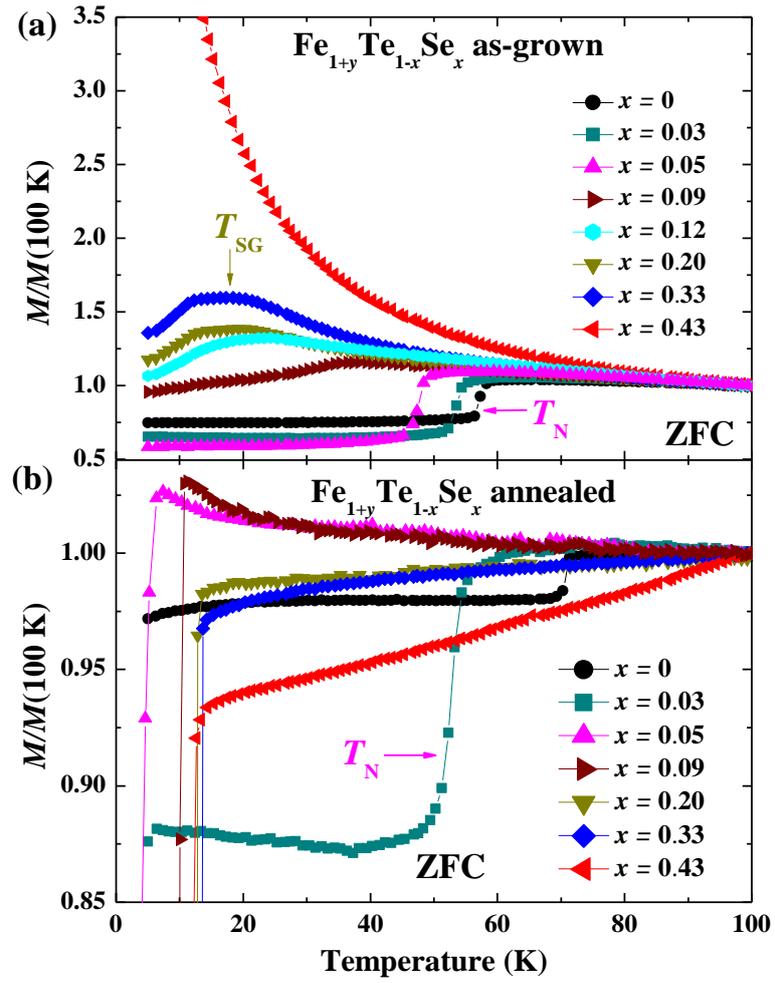

Figure 3

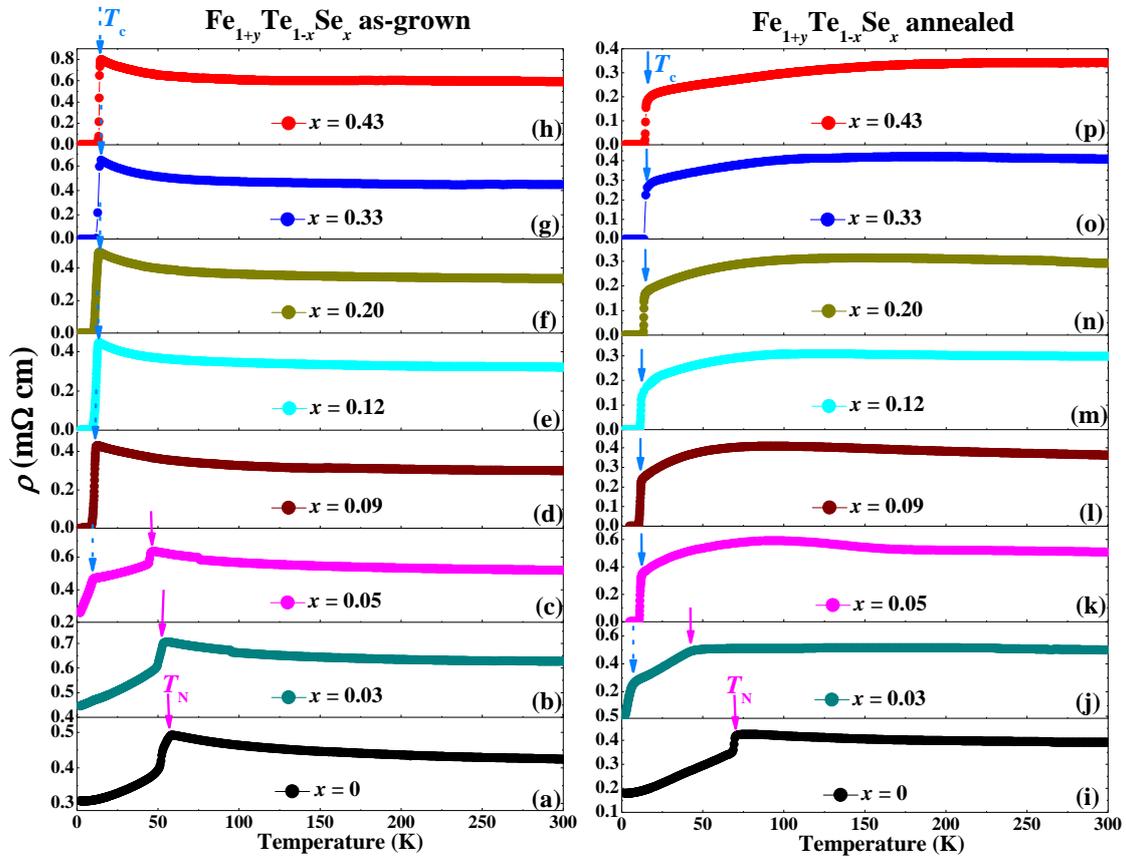

Figure 4

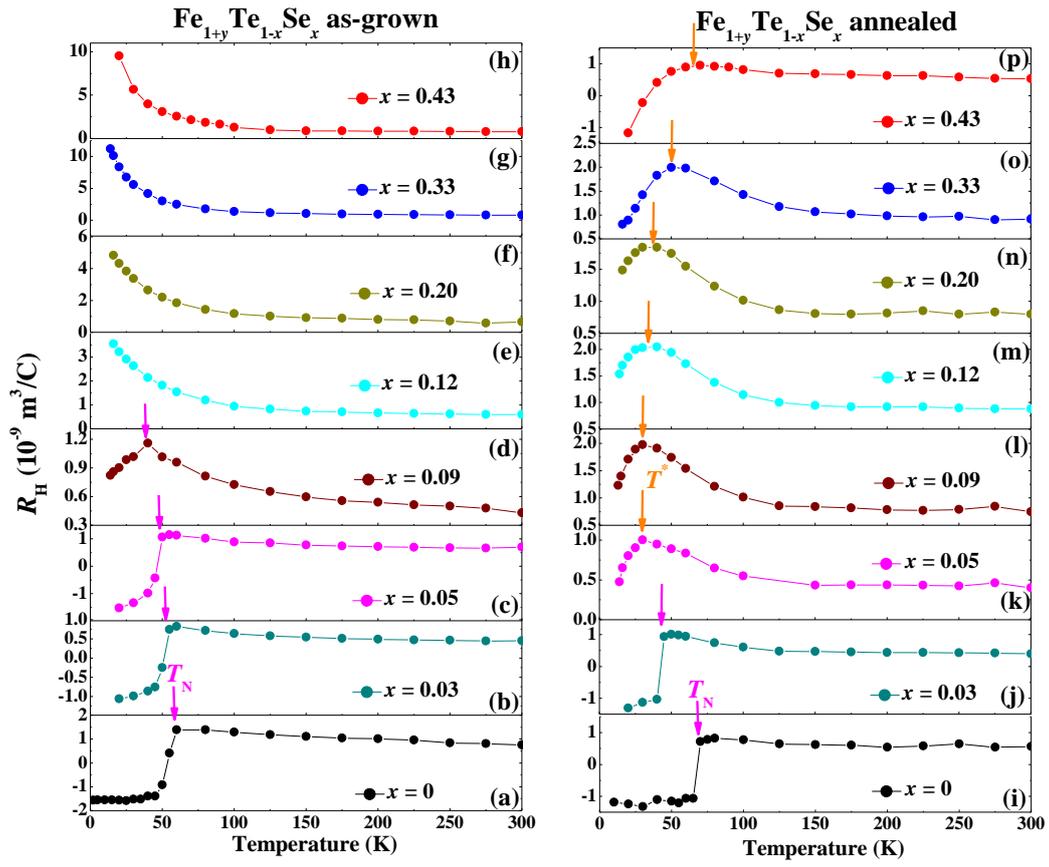

**Figure 5**

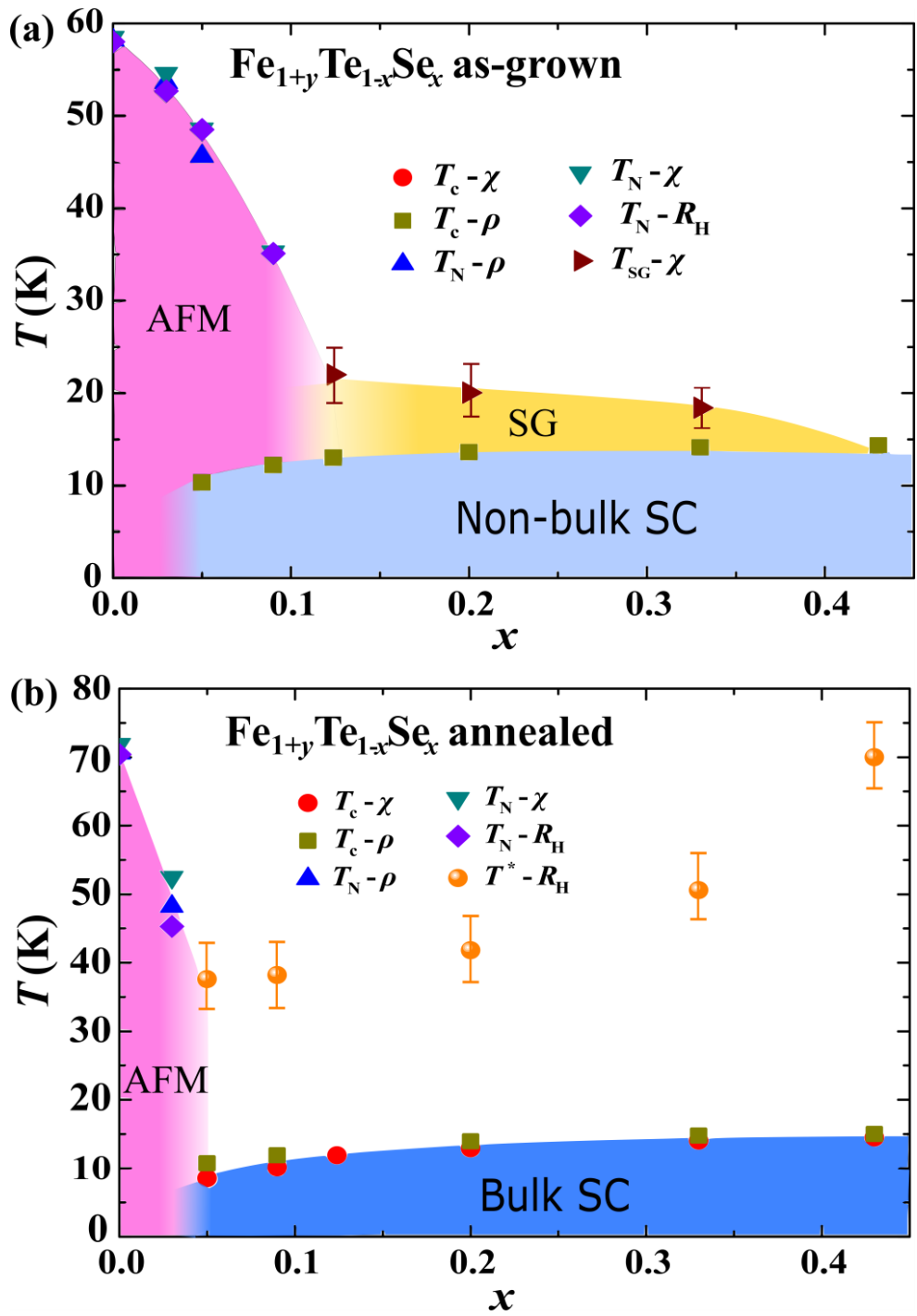

Figure 6